\documentclass{aa}
\usepackage{graphicx}
\begin{document}
\title{Local stability of self-gravitating fluid disks made of two components in relative motion}
\titlerunning{Local stability of self-gravitating fluid disks}

   \author{G. Bertin
          \inst{1}
          \and
          A. Cava\inst{1,2}
          }

   \offprints{G. Bertin}

   \institute{Universit\`a degli Studi di Milano, Dipartimento di Fisica, Via Celoria 16, 20133 Milano, Italy\\
              \email{giuseppe.bertin@unimi.it}
         \and
             INAF\,-\,Osservatorio Astronomico di Padova, Vicolo dell'Osservatorio 5, 35122 Padova, Italy\\
             \email{antonio.cava@oapd.inaf.it}
             }

   \date{Received; accepted}


  \abstract
   {We consider a simple self-gravitating
   disk, made of two fluid components characterized by different
   effective thermal speeds and interacting with one another only
   through gravity; two-component models of this type
   have often been considered in order to estimate the
   impact of the cold interstellar medium on gravitational instabilities
   in star-dominated galaxy disks.}
   {This simple model allows us to produce a unified description of instabilities
   in non-viscous self-gravitating disks, some originating from Jeans collapse, and others from
   the relative motion between the two components. In particular, the model suggests
   that the small streaming velocity between the two components associated with the
   so-called asymmetric drift may be the origin of instability for
   suitable non-axisymmetric perturbations.
   }
   {The result is obtained by
   examining the properties of a local, linear dispersion relation for
   tightly wound density waves in such two-component model. The parameters
   characterizing the equilibrium model and the related dispersion relation
   allow us to recover as natural limits the cases, known in the literature,
   in which the relative drift between the two components is ignored.}
   {Dynamically, the instability is similar to (although gentler than)
   that known to affect counter-rotating disks. However, in contrast to
   the instability induced by counter-rotation,
   which is a relatively rare phenomenon, the mechanism
   discussed in this paper is likely to be rather common in
   nature.}
   {We briefly indicate some consequences of the instability
   on the evolution of galaxy disks and possible applications to other
   astrophysical systems, in particular to protostellar disks and
   accretion disks.}

   \keywords{galaxies: kinematics and dynamics -- instabilities
               }

   \maketitle
%

\section{Introduction}

As clearly noted in investigations of the problem of spiral
structure in galaxies, the dynamics of galaxy disks depends
significantly on the mutual interaction between gas and stars. For
the general issues related to the properties of waves and
instabilities, such gas-star interactions have often been studied
by referring to a very simple, idealized two-component model in
which one distinguishes a ``hot" component (meant to represent
mostly the older stellar disk) from a ``cold" component
(representative of atomic and molecular hydrogen and young stars)
interacting with each other only through gravity. In practice,
calculations have been carried out by treating one component by
means of the equations of stellar dynamics and the other by fluid
equations, or, for simplicity, by treating both components by
means of simple barotropic fluid models characterized by different
effective thermal speeds; other investigations have implemented
these concepts by means of suitable N-body simulations. It has
thus been ascertained that even small amounts of gas can have
significant influence on the Jeans stability of galaxy disks
(e.g., see Lin \& Shu \cite{lin66}; Lynden-Bell \cite{lyn67};
Graham \cite{gra67}; Miller, Prendergast \& Quirk \cite{mil70};
Quirk \cite{qui71}; Kato \cite{kat72}; Jog \& Solomon
\cite{jog84a}, \cite{jog84b}; Sellwood \& Carlberg \cite{sel84};
Bertin \& Romeo \cite{ber88}, hereafter BR88).

Similar models have also been used to examine situations in which
the two components are in relative motion with respect to one
another. Among the earliest investigations, we may recall those by
Sweet (\cite{swe63}), Lynden-Bell (\cite{lyn67}), Wilson \&
Lynden-Bell (unpublished; as quoted in Lynden-Bell \cite{lyn67}),
and Marochnik \& Suchkov (\cite{mar69,mar74}). Often in such early
studies the ``Population II" stars are treated as a non-rotating
component, in a kinetic model, and, for simplicity, the geometry
adopted is that of a homogeneous three-dimensional system (see
also the extensive discussion of the interaction between gas and
stars in the monograph by Saslaw \cite{sas85}). Later on, similar
studies were further motivated by the issue of disk-halo
interaction (Mark \cite{mar76}); so, in the latter investigations
one component represents the rotating disk and the other the
non-rotating or slowly rotating halo material. These analyses then
brought out the possibility of a kind of two-stream or
Kelvin-Helmholtz instability, thus beyond Jeans instability, with
applications to the amplification of bending waves in relation to
the problem of galaxy warps (Bertin \& Mark \cite{ber80}; Bertin
\& Casertano \cite{ber82}; revisited by Nelson \& Tremaine
\cite{nel95}). With similar tools, others have explored the
possibility of interactions between different gaseous components
in galaxies (see Nelson \cite{nel76}; Waxman
\cite{wax79a,wax79b}).

More recently, renewed attention has been drawn to the study of
instabilities excited in disks in relative motion (Sellwood \&
Merritt \cite{sel94}; Lovelace, Jore \& Haynes \cite{lov97},
hereafter LJH97; Hunter, Whitaker \& Lovelace \cite{hun97}; Comins
et al. \cite{com97}; Mandica \cite{man01}) by the discovery of the
curious phenomenon of counter-rotation, noted in a number of disk
galaxies (Rubin, Graham \& Kenney \cite{rub92}; Rix et al.
\cite{rix92}; Merrifield \& Kuijken \cite{mer94}; Bertola et al.
\cite{ber96}; Jore, Broeils \& Haynes \cite{jor96}; Prada et al.
\cite{pra96}; Thakar et al. \cite{tha97}; Bettoni et al.
\cite{bet01}).

Counter-rotation appears to be a relatively rare, and possibly
short-lived phenomenon. Thus, such relative motion between two
disk components is likely to be only rarely involved in the
dynamics of galaxies. In contrast, in this paper we focus our
attention on a potential source of instability that might be
ubiquitous: because of the so-called asymmetric drift associated
with the finite velocity dispersion of the hot component (e.g.,
see Bottema \cite{bot93}), any disk of stars and gas would in
general be characterized by small, but finite relative motion
between the two components.

In this paper we will show that, in principle, the small velocity
difference between two components associated with the asymmetric
drift is a source of instabilities in the disk. This result will
be obtained by inspection of a local, linear dispersion relation
for tightly wound density waves derived from a simple two-fluid
model in Sect.~2. In Sect.~3 we will study the conditions for the
onset of the asymmetric-drift instability and its main
characteristics. Finally, in Sect.~4 we will describe possible
applications to different astrophysical contexts and draw the main
conclusions of our paper. Given the fact that such
asymmetric-drift velocity is small, for applications to galaxy
disks, this paper is meant to offer an investigation preliminary
to a fluid-kinetic analysis that is planned for the future.

While this work was being completed, we were reminded of an
interesting study made in the context of the dynamics of
protostellar disks (Weidenschilling \cite{wei77}), about the
viscous drag exerted by the gas of the protostellar disk on the
population of planetesimals, which presents significant analogies
with the mechanism addressed in this paper. We will briefly
comment on this point in the conclusion section.

\section{The fluid model and the relevant dispersion relation}

In a rotating self-gravitating axisymmetric fluid disk at
equilibrium the radial gravitational force is balanced by
rotation, with a contribution from the pressure gradient:

\begin{equation}
    \label{omega}
\Omega^2 = \frac{1}{r\sigma}\frac{d p}{d
r}+\frac{1}{r}\frac{d\Phi}{d r}~,
\end{equation}

\noindent where $\Omega,~\sigma$, and $p$ are the equilibrium
angular velocity, surface density, and pressure of the thin disk,
and $\Phi$ is the total gravitational potential. For cool disks,
the pressure gradient is small and generally neglected. Therefore,
many studies of the dynamics of such disks simply refer to the
angular frequency as defined by the relation $r \Omega_0^2 \equiv
d\Phi/dr$ and to the related epicyclic frequency defined by
$\kappa_0^2 \equiv 4 \Omega_0^2 [1 + (1/2) d \ln{\Omega_0/d
\ln{r}}]$.

We now consider an infinitesimally thin, self-gravitating,
two-component disk. In general, the two components are
characterized by different equivalent acoustic speeds $c_{i}$ and
surface densities $\sigma_{i}$. We label the two components by the
subscripts $c$ and $h$ (in relation to their equivalent acoustic
speeds, so that $c_c < c_h$). Similarly to other previous
stability analyses, we assume that the only interaction between
the components occurs via the gravitational field. In contrast
with previous analyses that  focused on the Jeans instability and
assigned the same equilibrium angular velocity $\Omega_0$ to the
two components (e.g., see BR88), in this paper we allow for the
presence of relative motion between the two components, so that in
general $\Omega_c \neq \Omega_h$, and, in turn, for a difference
in the associated epicyclic frequencies $\kappa_{c} \neq
\kappa_h$. The special limiting case in which $c_c \longrightarrow
c_h$ and $\Omega_c \longrightarrow - \Omega_h$ thus corresponds to
the case of counter-rotating components, a situation already
discussed in the literature, as mentioned in the Introduction. We
argue that, because of the different size of the relevant
asymmetric drift dictated by Eq.~(\ref{omega}) the angular
velocities of the two components would in general be (although
slightly) different even when the two components are not
counter-rotating.

\subsection{Applicability of the two-fluid model}

The two interpenetrated non-viscous fluids studied in this paper
are to be taken as the simplest model in which we can include
several key factors (temperature, density, and velocity
differences, within the non-trivial geometry of an inhomogenous
thin layer) that determine the stability of a two-component
self-gravitating disk. Some important features, such as finite
thickness, present in real self-gravitating disks are not
included. Therefore, for any specific application (for example, to
the dynamics of galaxy disks) it will be necessary to check to
what extent the basic approximations implicit in this model are
justified and, if necessary, to extend the calculations to a more
realistic (and thus more complicated) model. On the other hand,
studying this simple model has the advantage of leading to a
unified picture of a variety of instabilities that may affect a
self-gravitating disk.

One particular point of concern that will emerge from the
following analysis is related to the applicability of the
two-fluid model to cases in which the relative speed between the
two components is small. In fact, it is well known that, for a
kinetic system made of two components in relative streaming, the
fluid description becomes inadequate when the relative speed is
small (e.g., see Lynden-Bell \cite{lyn67}, Krall \& Trievelpiece
\cite{kra73}). The instability noted in a two-fluid description
may remain, but is usually modified because the underlying
mechanism of instability is associated with an inverse Landau
damping (i.e., the effects of wave-particle resonances) that only
a kinetic description can properly describe. Thus, in view of
applications to a disk made of stars and gas (a system that is
neither fully kinetic nor fully fluid), the present study should
be completed by investigating the mechanisms in a mixed
fluid-kinetic model; this extension will be addressed in a
separate paper.

\subsection{Local dispersion relation}

We now consider tightly wound linear density perturbations of the
form $\sigma_{1i}=\hat{\sigma}_{1i} \exp[i(-\omega t  + m\theta +
\int^r k dr)]$ under the WKB ordering $m/(r|k|) = O(\epsilon)$,
with the epicyclic expansion parameter defined as $\epsilon =
c_h/(r\kappa_h)$. We will take the azimuthal number $m$ to be
positive, so that trailing disturbances are characterized by
radial wave number $k > 0$. To lowest order, the density response
for each component (resulting from the relevant continuity and
Euler equations) is given by

\begin{equation}
    \label{dens}
    \sigma_{1i}=\sigma_{i}
    \frac{k^2\Phi_1}{(\omega-m\Omega_{i})^2-\kappa_{i}^2-c_{i}^2k^2}~.
\end{equation}

\noindent Each component experiences the joint perturbed
gravitational potential $\Phi_1$, which is calculated from the
Poisson equation

\begin{equation}
    \label{poi}
    -|k|\Phi_1=2\pi G(\sigma_{1c}+\sigma_{1h})~.
\end{equation}

\noindent By combining the above equations to eliminate $\Phi_1$
(for a complete derivation see Cava \cite{cav04}), we obtain the
dispersion relation, which we write in dimensionless form as

\begin{equation}\label{disprel}
\nu^4-2\eta\nu^3 - A^{(2)}\nu^2 + 2 \eta A^{(1)}\nu + A^{(0)} =
0~.
\end{equation}

\noindent Here the relevant coefficients $A^{(n)} = A^{(n)}
(\hat{|k|}; Q_h^2, \alpha, \beta, \Delta)$ for $n \neq 1$ and
$A^{(1)} = A^{(1)}(\hat{|k|};Q_h^2)$ are defined as
\begin{equation}
A^{(2)} = 1 + A^{(1)} + \Delta - \alpha \hat{|k|} +
\frac{1}{4}Q_h^2\beta\hat{|k|}^2~,
\end{equation}

\begin{equation}
A^{(1)} = 1-\hat{|k|}+\frac{1}{4}Q_h^2 \hat{|k|}^2~,
\end{equation}

\begin{eqnarray}\label{a0}
A^{(0)} = (1 + \Delta) A^{(1)} - \alpha \hat{|k|}
+\frac{1}{4}Q_h^2 \beta \hat{|k|}^2-\\
\nonumber\;\frac{1}{4}Q_h^2(\alpha+\beta)\hat{|k|}^3+
\frac{1}{16}\beta Q_h^4 \hat{|k|}^4~.
\end{eqnarray}

\noindent Here $\hat k = 2\pi G\sigma_{h}k/\kappa_{h}^2$ and $\nu
= (\omega - m\Omega_{h})/\kappa_{h}$ are the dimensionless radial
wavenumber and the dimensionless Doppler-shifted frequency,
respectively. The quantity $Q_h = c_{h}\kappa_{h}/(\pi G\sigma_h)$
is the axisymmetric stability parameter relative to the hot
component.

Thus, with respect to the one-component stability analysis (which
is uniquely controlled by the axisymmetric stability parameter),
the dispersion relation depends on four additional dimensionless
parameters $\alpha,\beta,\Delta$, and $\eta$. For the first two
(positive definite) parameters we follow the definition of BR88,
i.e. $\alpha = \sigma_{c}/\sigma_{h}$ and $\beta = c_{c}^2/c_{h}^2
< 1$ are the relative density and ``temperature" ratios. The two
new parameters are defined in terms of the angular velocity ratio
$\hat\delta \equiv \Omega_{c}/\Omega_{h}$ (which, for simplicity,
was taken to be constant in the derivation of Eq.~(\ref{disprel}))
as $\eta = m(\hat\delta-1)(\Omega_{h}/\kappa_{h})$ and $\Delta =
\hat\delta^2 - \eta^2 - 1$; therefore, in the limit $\Omega_c
\longrightarrow \Omega_h$ we have $\Delta \longrightarrow 0$ and
$\eta \longrightarrow 0$. Note that $\eta$ vanishes for
axisymmetric perturbations ($m = 0$) even when the two components
are in relative motion ($\hat\delta \neq 1$). Since we ascribe the
difference in angular velocity to the different weight of the
pressure term in Eq.~(\ref{omega}), we expect the cooler component
to rotate faster, so that in general $|\hat\delta| > 1$.

\subsection{Recovery of known limiting cases}

For $\hat\delta = 1$ the two components are corotating exactly,
which is the case considered by BR88, while for $\hat\delta = -1$
the components are counter-rotating, which corresponds to the case
considered by LJH97 (who, at variance with the present two-fluid
description, considered a mixed fluid-kinetic model and a
kinetic-kinetic model).

The one-component dispersion relation $\nu^2 = A^{(1)}(\hat{|k|};
 Q_h^2)$ is recovered by taking the limit $\beta \longrightarrow 1$,
$\hat\delta \longrightarrow 1$ (so that $\eta \longrightarrow 0$
and $\Delta \longrightarrow 0$), and by letting $\alpha$ become
vanishingly small; the second, odd branch (see Sect.~2.2 in BR88)
of the dispersion relation then reduces to the condition $\nu^2 =
1 + (1/4)Q_h^2 \hat{|k|}^2$ for modified sound waves in the ``gas
tracer". For a two-component system the limit of exactly
corotating fluids is obtained by letting $\eta = 0$ and $\Delta =
0$ so that the dispersion relation reduces to $\nu^4 -
A_0^{(2)}\nu^2 + A_0^{(0)} = 0$, with $A_0^{(0)}(\hat{|k|}; Q_h^2,
\alpha, \beta) = A^{(1)} - \alpha \hat{|k|} + (1/4)Q_h^2 \beta
\hat{|k|}^2 -(1/4)Q_h^2(\alpha+\beta)\hat{|k|}^3 + (1/16)\beta
Q_h^4 \hat{|k|}^4$; thus the marginal stability curve at the basis
of the analysis of BR88 (see their Eq.~(14)) is recovered by
setting $A_0^{(0)} = 0$.

\section{The asymmetric-drift instability}

\subsection{\label{ss11}Axisymmetric disturbances: reduction to
standard Jeans stability conditions}

We first consider the case of axisymmetric disturbances, that is
the case of $\eta = 0$, with $\Delta = \Delta_0 \equiv
\hat\delta^2 - 1 > 0$. Here the dispersion relation reduces to
$\nu^4 - A^{(2)}\nu^2 + A^{(0)} = 0$. Therefore, the marginal
stability condition is given by $A^{(0)} = 0$. From Eq.~(\ref{a0})
it is clear that, even if $\Delta \neq 0$, the condition is the
same as for the $\Delta = 0$ case, provided the relevant
parameters $\alpha$ and $\beta$ are suitably rescaled. In fact, if
we define $\alpha' = \alpha/(1 + \Delta_0)$ and $\beta' = \beta/(1
+ \Delta_0)$, we find that the condition $A^{(0)}(\hat{|k|};
Q_h^2, \alpha, \beta, \Delta_0) = 0$ is equivalent to the
condition $A_0^{(0)}(\hat{|k|}; Q_h^2, \alpha', \beta') = 0$.
Therefore, for axisymmetric disturbances the stability condition
is the same as for the standard Jeans instability in two-component
disks. In particular, local stability at {\it all} wavelengths is
guaranteed when $Q_h \geq \bar{Q}(\alpha',\beta')$, which replaces
the local axisymmetric stability condition for one-component fluid
disks $Q \geq 1$; the function $\bar{Q} = \bar{Q}(\alpha,\beta)$
is illustrated in Fig.\,\ref{fig5} of BR88. Note that, in general,
for $\Delta_0
>0$ we have $\bar{Q}(\alpha',\beta') < \bar{Q}(\alpha,\beta)$,
which corresponds to a {\it stabilizing} effect on $m=0$
perturbations. We should also point out that the rescaling of
$\alpha$ and $\beta$ that has allowed us to make use of the
marginal stability condition described in BR88 does not correspond
to exactly mapping the roots of the dispersion relation with
$\hat\delta >1$ into those of the case with $\hat\delta = 1$.

\subsection{Instability for non-axisymmetric perturbations at small values of $\eta$}

We now refer to the case when the relative motion between the two
components is small, that is $\eta \ll 1$, with $\Delta = \Delta_0
- \eta^2 < \Delta_0$, and consider an equilibrium case that is
marginally stable with respect to the Jeans instability, so that
$Q_h = \bar{Q}( \alpha/(1 + \Delta_0),\beta/(1 + \Delta_0))$. The
``most dangerous perturbations" are those with wavenumber close to
$\bar{|k|}$, defined by the relation $A_0^{(0)}(\bar{|k|};
\bar{Q}^2, \alpha/(1 + \Delta_0),\beta/(1 + \Delta_0)) = 0$; close
to $\bar{|k|}$, the quantity $A^{(0)}$ is expected to be {\it
negative} and small, because $\Delta < \Delta_0$. In addition, we
expect $A^{(1)}>0$ (the last relation follows from the fact that
the presence of a second component is known to have a
destabilizing effect on axisymmetric perturbations, i.e.
$\bar{Q}(\alpha',\beta')
> 1$). As to the value of the frequency for such conditions close to
marginal stability, we expect it to be also small, and thus we can
approximate the relevant dispersion relation as a quadratic

\begin{equation}
- A^{(2)}\nu^2 + 2\eta A^{(1)}\nu + A^{(0)} \sim 0~.\label{drred}
\end{equation}

\noindent Therefore, in the wavenumber range close to $\bar{|k|}$
in which

\begin{equation}
\eta^2 (A^{(1)})^2 + A^{(2)} A^{(0)} < 0 \label{cond_a}
\end{equation}

\noindent we find a pair of complex-conjugate solutions one of
which corresponds to instability, with dimensionless growth rate
$\nu_I = O(\eta)$. The real part of the frequency for these
complex solutions is given by

\begin{equation}
\nu_R \sim \eta \frac{A^{(1)}}{A^{(2)}} < \eta~,
\end{equation}

\noindent which, as anticipated, corresponds to a pattern speed
$\Omega_p \equiv \omega_R/m$ in the range $\Omega_h < \Omega_p <
\Omega_c$.

\begin{figure}
\includegraphics{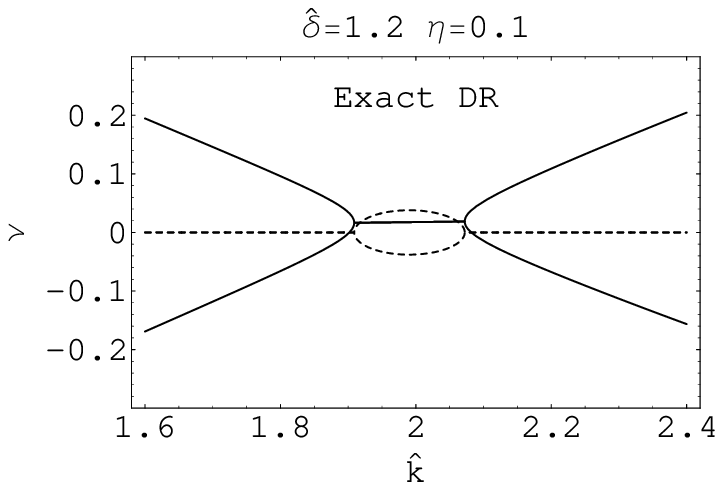}
\includegraphics{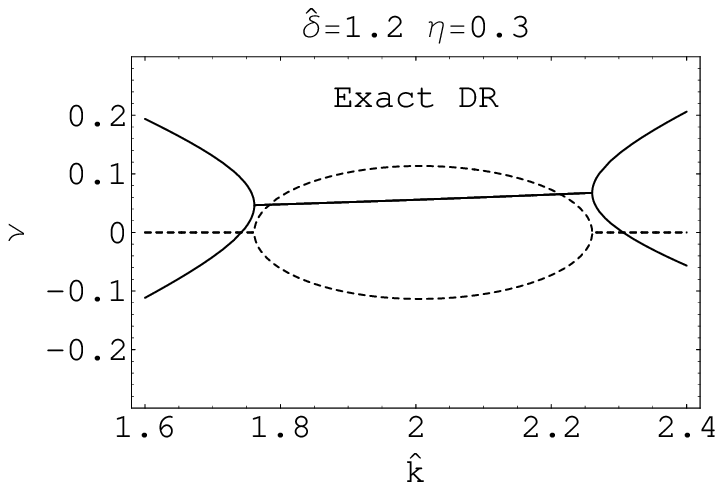}
\includegraphics{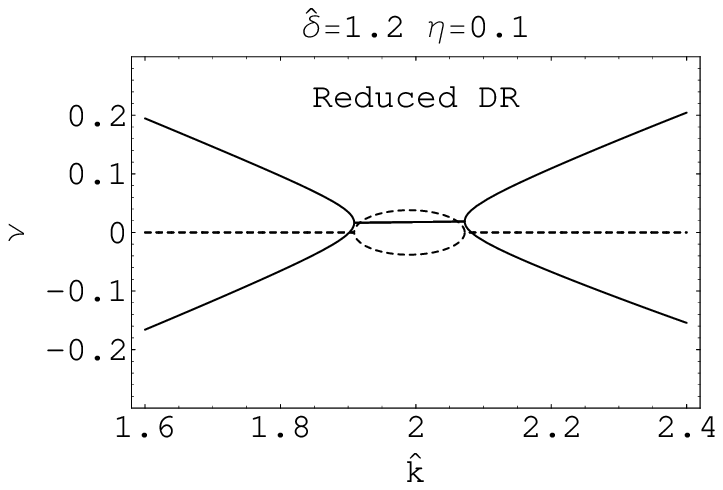}
\includegraphics{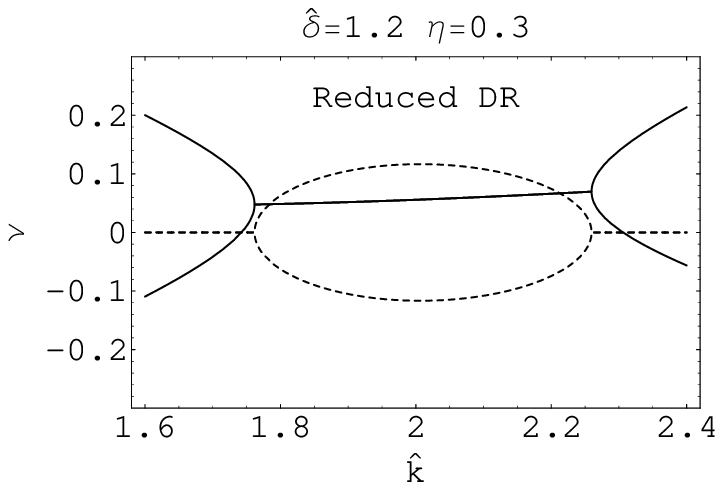} \caption{\small{The
dispersion relation in the vicinity of marginal stability from the
exact (Eq.\,\ref{disprel}; top frames) and from the ``reduced"
(Eq.\,\ref{drred}; bottom frames) expression, for two cases
characterized by $\hat\delta = 1.2$, $\eta = 0.1$ and $\hat\delta =
1.2$, $\eta = 0.3$. In each frame the real (solid lines) and
imaginary (dashed lines) part of $\nu$ are given as a function of
the dimensionless wavenumber $\hat{k}$. The other parameters have
been set to the following values: $\alpha = 0.1$, $\beta = 0.1$,
$Q_h =\bar{Q}(\alpha/\hat\delta^2,\beta/\hat\delta^2)\approx 1.14$.
The ``reduced" dispersion relation turns out to be accurate even for
$\eta = 0.3$. }} \label{fig1}
    \end{figure}

\begin{figure}
\includegraphics{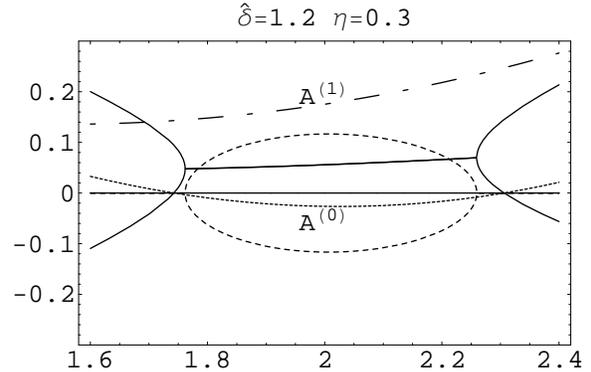}
\caption{\small{Illustration of the generation of the unstable
roots. Superposed to the dispersion relation shown in the second
frame of Fig.\,\ref{fig1} (for the case $\hat\delta = 1.2$, $\eta =
0.3$, with $\alpha = 0.1$, $\beta = 0.1$, $Q_h \approx 1.14$) we
plot the coefficients $A^{(0)}$ (dotted curve) and $A^{(1)}$
(dash-dotted curve) for values of $\hat{k}$ close to $\bar{k}$. This
diagram is given in support of the arguments that lead to the
condition (Eq.\,\ref{cond_a}) for the onset of the instability.}}
\label{fig2}
    \end{figure}

These facts are illustrated in Fig.\,\ref{fig1} , where we compare
the dispersion relation in the vicinity of marginal stability from
the exact (Eq. 4) and from the "reduced" (Eq. 8) expression,
showing that the approximate quadratic relation can be accurate
even for relatively large values of $\eta$. In Fig.\,\ref{fig2} we
also illustrate the behavior of the coefficients $A^{(0)}$ and
$A^{(1)}$, in line with the arguments given above to derive the
onset of the instability analytically.

\subsection{Non-axisymmetric perturbations studied in the general case}

At this point we can move on to study the general case, when the new
parameters $\Delta$ (or $\hat\delta $) and $\eta$ that appear in the
dispersion relation cannot be taken to be small. As described
previously, we have in mind two different kinds of situations of
astrophysical interest. The case of counter-rotation is associated
with $\hat\delta \approx -1$ so that we expect $\eta$ and $\Delta$
to have finite values. As noted earlier in the Introduction, this
case has already been addressed in the literature. The second case,
where the velocity difference of the two components, being generated
by the asymmetric drift, is small, has been dealt with in the
previous subsection; here, in order to consider $\eta = O(1)$, we
should take $m$ to be large. In this new regime of open waves the
basic dispersion relation Eq.~(\ref{disprel}) would be modified and
there would exist additional destabilizing mechanisms (see Bertin et
al. \cite{ber89}; note the threshold value indicated by their
Eq.~(5.8), which, for a flat rotation curve, reduces to $Q^2 = 3$).
For this, a consistent asymptotic analysis might be developed in an
approach similar to the one followed in this paper, but for
simplicity we omit such analysis in this paper.

\begin{figure}
\includegraphics[scale=.9]{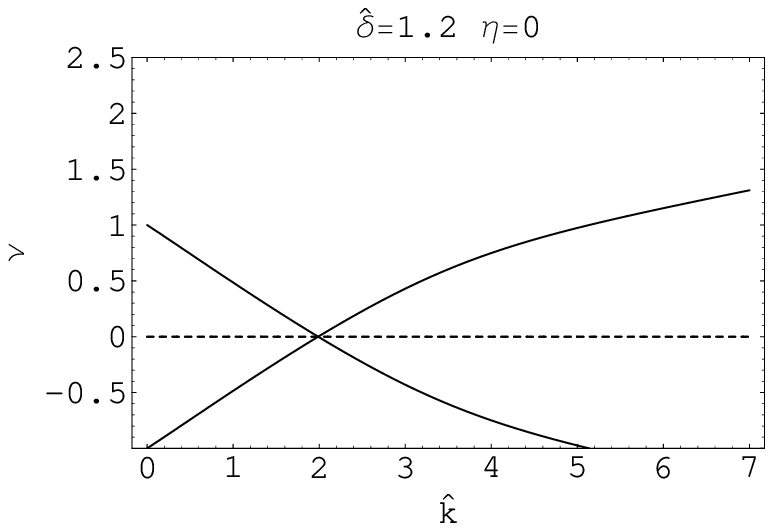}
\includegraphics[scale=.9]{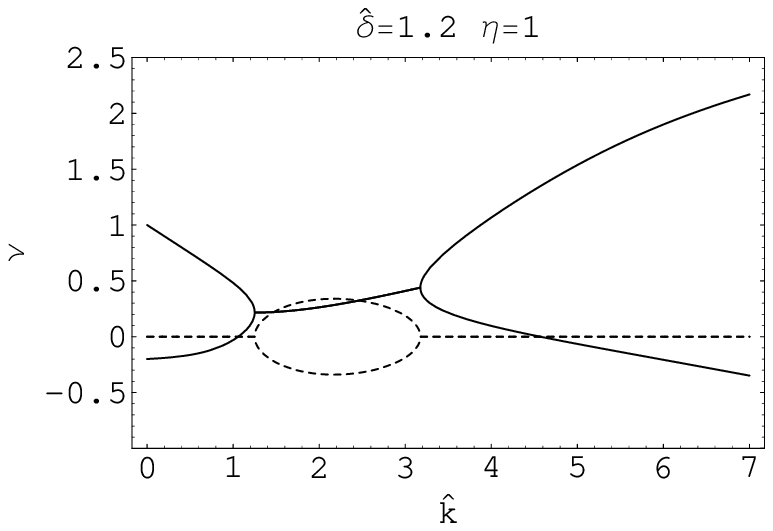}
\includegraphics[scale=.9]{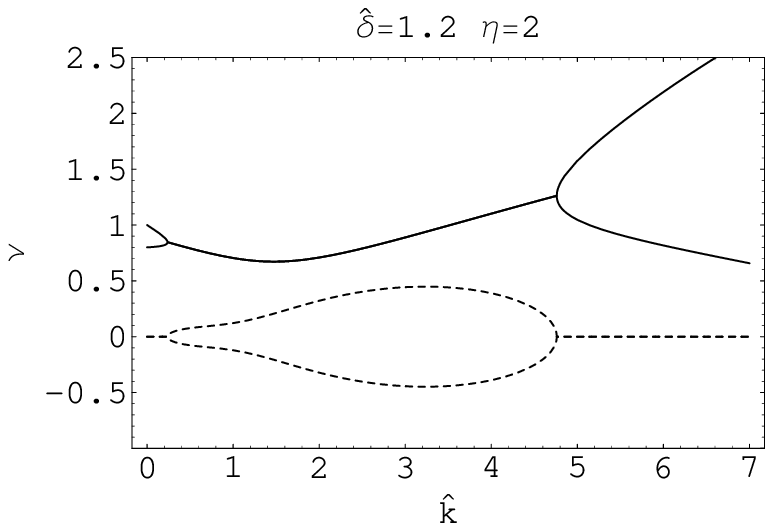}
\includegraphics[scale=.9]{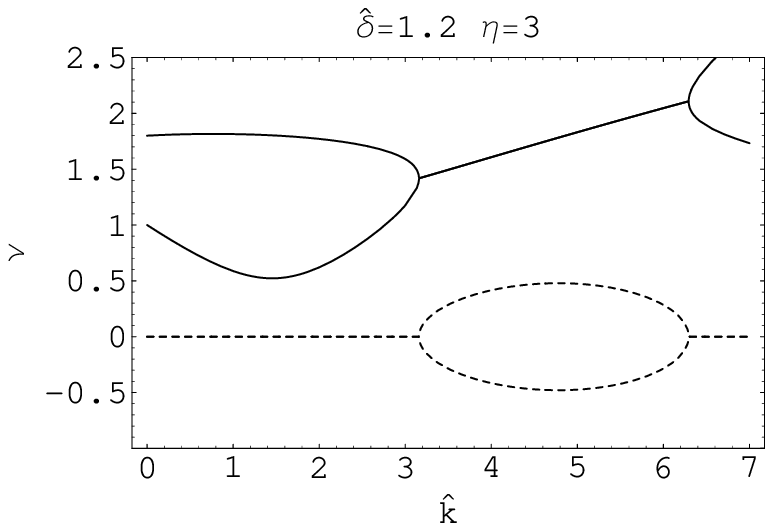}
\caption{\small{The dispersion relation for varying $\eta$. In each
frame the real (solid lines) and imaginary (dashed lines) part of
$\nu$ are given as a function of the dimensionless wavenumber
$\hat{k}$, for $\eta=0$, $\eta = 1$, $\eta = 2$, $\eta = 3$ (from
top to bottom). The other parameters have been set to the following
values: $\hat\delta = 1.2$, $\alpha = 0.1$, $\beta = 0.1$, $Q_h =
\bar{Q}(\alpha/\hat\delta^2,\beta/\hat\delta^2) \approx 1.14$.}
\label{fig3}}
    \end{figure}

\begin{figure}
\includegraphics[scale=.9]{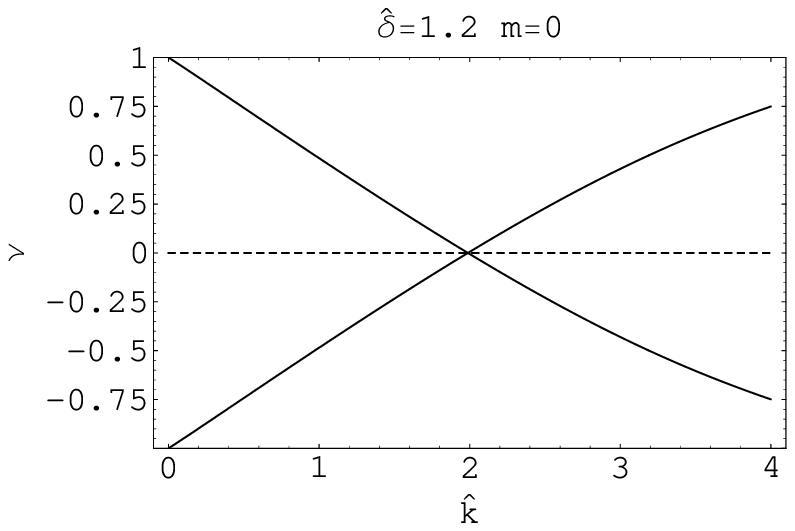}
\includegraphics[scale=.9]{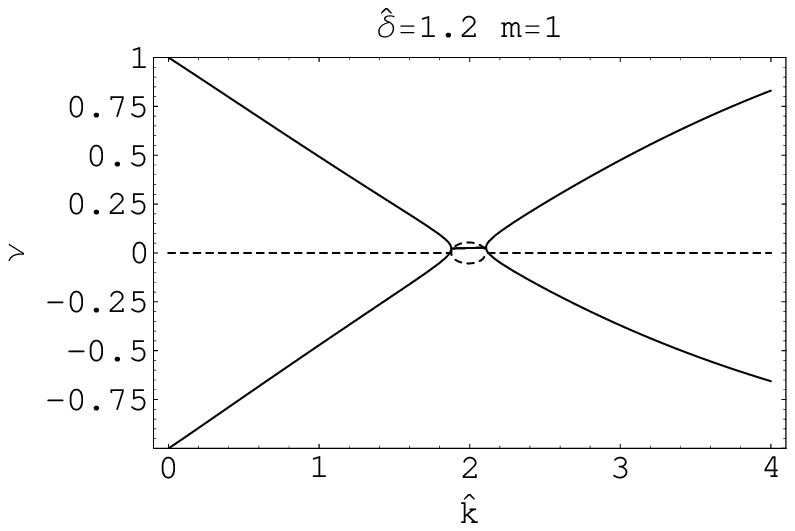}
\includegraphics[scale=.9]{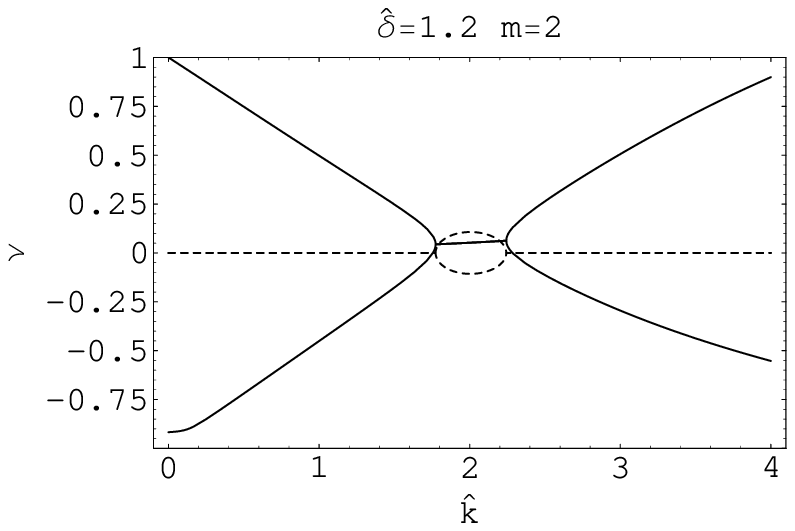}
\includegraphics[scale=.9]{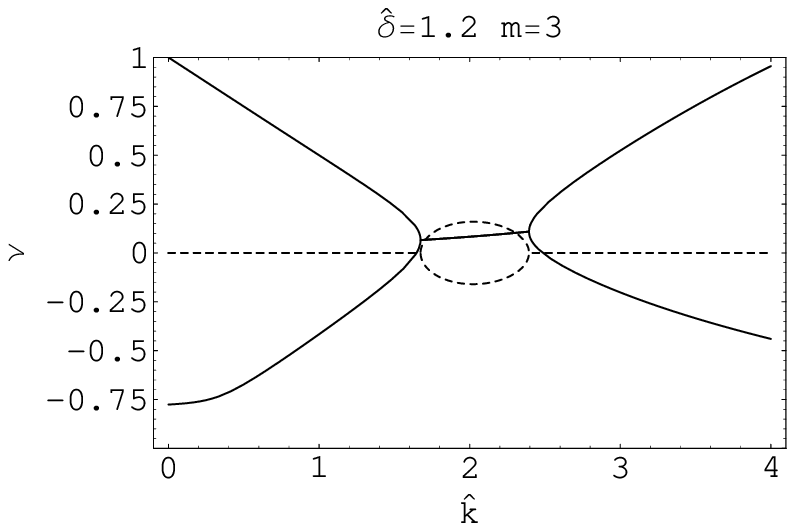}
\caption{\small{The dispersion relation for varying m. In each frame
the real (solid lines) and imaginary (dashed lines) part of $\nu$
are given as a function of the dimensionless wavenumber $\hat{k}$,
for $m = 0$, $m = 1$, $m = 2$, $m = 3$ (from top to bottom). The
other parameters have been set to the following values: $\hat\delta
= 1.2$, $\alpha = 0.1$, $\beta = 0.1$, $Q_h =
\bar{Q}(\alpha/\hat\delta^2,\beta/\hat\delta^2) \approx 1.14$, with
the additional condition of flat rotation curve
$\Omega_{h}/\kappa_{h} =\sqrt{2}/2$.}} \label{fig4}
    \end{figure}

In Figs.\,\ref{fig3}-\ref{fig4} we illustrate the qualitative
behavior of the dispersion relation at fixed value of $\hat\delta$
for varying $\eta$ and for varying $m$ (with fixed
$\Omega_{h}/\kappa_{h}$); in each plot the dashed and solid lines
represent the imaginary and real part of the dimensionless
frequency $\nu$, respectively. Since, at variance with the case of
axisymmetric perturbations, the unstable root has a non-vanishing
real part of the frequency, we are in the presence of a {\it
convective} instability (see LJH97). Following the general
arguments used by LJH97, we can also confirm for this general case
of $\eta = O(1)$ that instability can develop when
$\nu_R(\nu_R-\eta)<0$, i.e. when $\Omega_h < \Omega_p < \Omega_c$.

In conclusion, the dispersion relation indicates the existence of
unstable roots with dimensionless growth-rate (imaginary part of
$\nu$) up to $0.1 - 0.2$ for small values of $m$, and formally up
to $0.4$ for higher values of $m$.

\subsection{Relevance to astrophysical applications}

As we will discuss further in the following Section, the main
context for astrophysical applications that we have in mind is that
of normal spiral galaxies.

\begin{figure}
\includegraphics{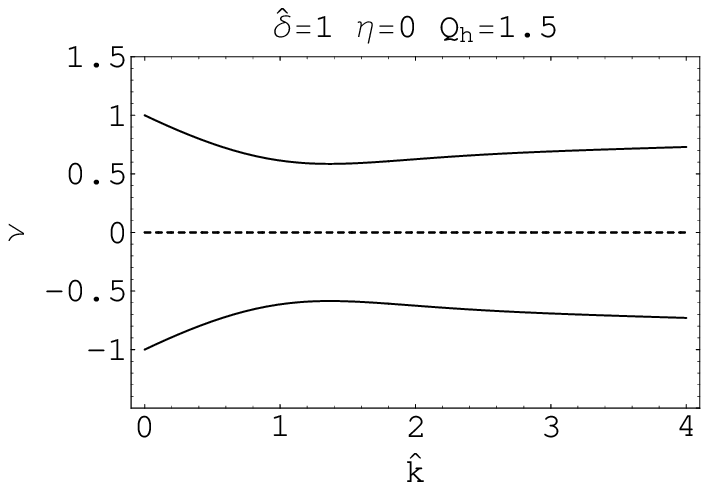}
\includegraphics{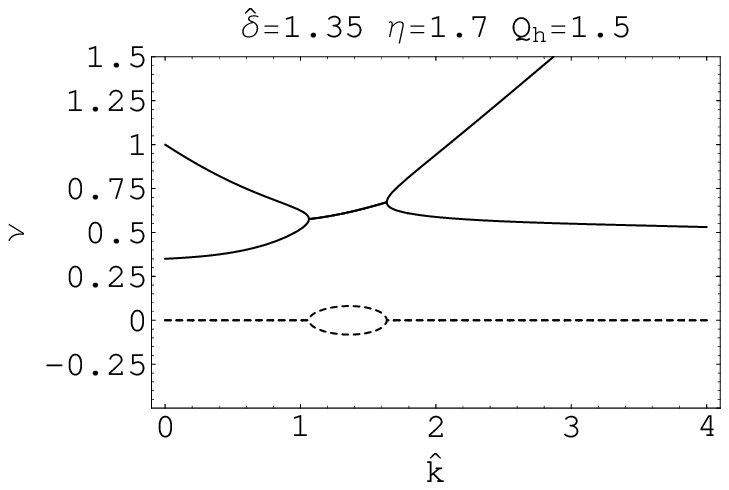}
\includegraphics{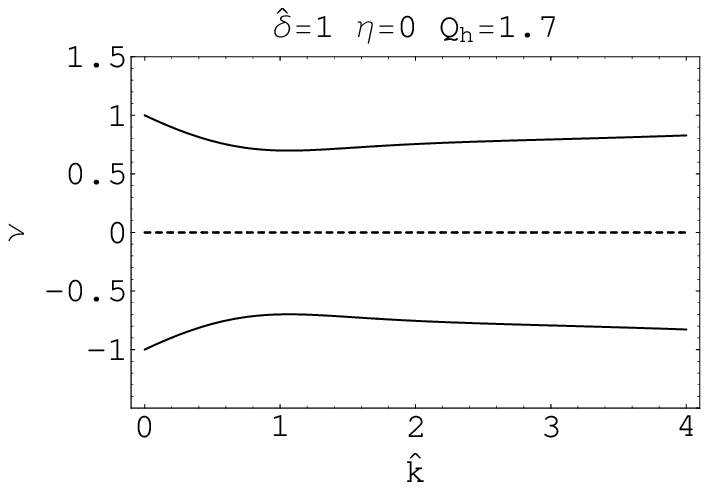}
\includegraphics{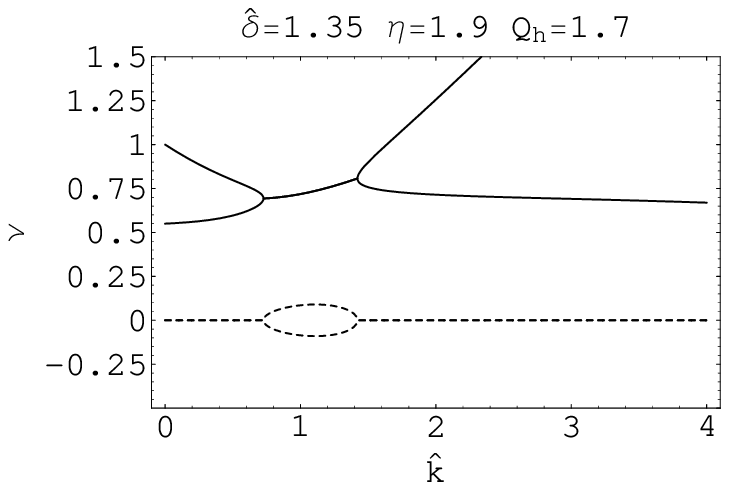}
\caption{\small{The asymmetric drift instability for a disk stable
with respect to axisymmetric instabilities, $Q_h>\bar
Q(\alpha,\beta)$. Here $\alpha = 0.1, \beta = 2.5\,10^{-2}$, values
suggested by conditions in the disk of the Milky Way. The vertical
axes in the second and fourth frames start from -0.50 to better
illustrate the imaginary part of the root of the dispersion relation
(Eq.\,\ref{disprel}).}} \label{fig5}
\end{figure}

The strategy that we have used has been to focus on a condition of
marginal stability with respect to axisymmetric perturbations in
the two-component disk ($Q_h = \bar{Q}(\alpha, \beta)$) in order
to demonstrate that the relative motion of the two components,
independently of axisymmetric Jeans-type instabilities, is
destabilizing. The instability that we have addressed can
obviously occur under more general conditions than $Q_h =
\bar{Q}(\alpha, \beta)$, as shown in Fig.\,\ref{fig5}; in
particular, instability can occur for $\alpha = 0.1$ and $\beta =
2.5\,10^{-2}$ (parameter values suggested by studies of the solar
neighborhood) even for $Q_h> 1.5$.

The asymmetric drift instability, being driven by the relative
motion of the two components, is expected to be ubiquitous and to
coexist (as mentioned in the first paragraph of $\S 3.3$) with
other sources of instability. As a result, the destabilizing
effect that we have discussed, even if present in realistic
simulations of two-component disks, may be difficult to single out
from other sources of instability and from a number of effects
related to the specific modeling used in each simulation (e.g.,
see Orlova et al. \cite{orl}; see also Noh et al. \cite{noh},
Korchagin \& Theis \cite{kor}).

Interestingly, the instability described in this paper is expected
to operate non-linearly to couple the two components (that is to
keep cold and hot components under physical conditions not too far
apart), counteracting the effects of the cooling processes present
in the cold gas, which favor the decoupling of the two components.
This makes the asymmetric drift instability participate in the
"thermostat" of self-regulation (necessary for grand design spiral
structure to be maintained; see Bertin \& Romeo \cite{ber88},
Bertin et al. \cite{ber89}). The natural emergence of
self-regulation has been demonstrated recently in numerical
experiments of the dynamics of self-gravitating protostellar disks
(Lodato \& Rice \cite{lod04}).

Another point that we wish to emphasize is that the study that we
have performed is not limited to ``small-scale instabilities" (the
complex unstable root shown in Figs.\,\ref{fig4} and \ref{fig5}
occurs at $\hat{k}$ of order unity, which does not correspond to
``small scale" wavelengths); in fact, this property distinguishes
the mechanism studied in this paper from the destabilizing effect on
axisymmetric stability associated with the presence of cold gas,
which indeed tends to act on the small scale.

\section{Discussion and conclusions}

In this paper we have shown that a self-gravitating thin disk made
of a hotter and a cooler component can be unstable (with respect
to non-axisymmetric density waves) because of the different
amounts of asymmetric drift associated with the two components,
even when the disk is stable in the sense of Jeans (with respect
to axisymmetric density waves). The unified analysis performed
here is based on a local dispersion relation from which we can
easily recover the limits of previous stability investigations. In
particular, we can recognize the two-component Jeans stability
conditions, as studied by BR88, and the instability of
counter-rotating disks, as discussed by LJH97 (in a technically
more complicated study). For the case of small relative velocity
between the two components, which is the focus of the present
paper, and of low values of the azimuthal wavenumber $m$, the
growth rate of the instability has been derived analytically. Here
we aimed at investigating the asymmetric drift instability
mechanism by itself and were not interested in assessing the
stability or instability of a disk under specific physical
conditions. For this latter purpose, it is well known that a
number of other factors (such as the role of thickness; e.g., see
Romeo \cite{rom92}) should be properly taken into account. The
collective effects realized in specific contexts will depend on
the conditions of the basic state and on the non-linear phenomena
that take place during evolution.

For applications to galaxy disks, the fluid-fluid analysis
presented here is preliminary to a more realistic analysis in
which the cold component is treated as a fluid, while the hotter
component is treated in terms of the collisionless Boltzmann
equation. Resonance effects, to be described by the equations of
stellar dynamics, are expected to occur when the velocity
difference between the two components is small, i.e. when it
becomes comparable to the velocity dispersion of the stars
associated with the hotter component; note that this condition
refers to the velocity dispersion in the {\it azimuthal
direction}, which is smaller by approximately a factor of
$\kappa/(2 \Omega)$ than the radial velocity dispersion, which is
responsible for the asymmetric drift. A proper description of
resonances will lead to technical complications well beyond those
addressed in the fluid-kinetic study of LJH97. Because of its
physical origin, the asymmetric drift velocity is {\it a priori}
bound to be relatively small, even for those galaxies (e.g.,
NGC~5064, NGC~3200, and NGC~2815) in which such velocity
difference has been found to be, in magnitude, rather large (in
the range of 50 to 100 km/s; see Vega Beltr\'{a}n et al.
\cite{veg01} and Pizzella et al. \cite{piz04}).

For galaxy disks, the instability studied in this paper is likely
to induce turbulence and heating in the cold component and to
contribute to the generation of a net inflow of material toward
the center. In principle, one might think of a possible connection
with the topic of extra-planar gas that is currently the subject
of several investigations; this is the discovery that many galaxy
disks, in addition to their ``normal" thin HI layer, possess an
``anomalous" thick HI component characterized by slower rotation
(e.g., for NGC~2403, see Fraternali et al. \cite {fra02}; for
UGC~7321, see Matthews \& Wood \cite{mat03}; for NGC~4559, see
Barbieri et al. \cite{bar05}). However, for these systems it is
likely that the two HI components do not interpenetrate, but
rather occupy different heights with respect to the equatorial
plane; in this case, if instability occurs, the mechanism might be
different from the one studied in this paper, and more in line
with processes of the type investigated by Waxman
(\cite{wax79a,wax79b}).

Finding a mechanism able to act as a source of turbulence in disks
brings us to consider possible applications to the long-standing
issue of finding an adequate explanation for the anomalous
viscosity that is thought to operate in accretion disks (see
Shakura \& Sunyaev \cite{sha73}). The present mechanism would
contribute in those cases in which the various physical processes
support the coexistence of two components at different levels of
rotation around the central object. As a kind of two-stream
instability, it might operate even for plasma disks.

This latter application naturally brings us to consider the case
of protostellar disks. In fact, the mechanism studied in this
paper appears to be the collective counterpart to the ``viscous
drag mechanism" (Weidenschilling \cite{wei77}) that initially
appeared to act against the picture of planet growth from
planetesimals. Recent simulations have shown that such drag may
actually have a beneficial role in the direction of planet
formation (see Rice et al. \cite{ric04}). This paper suggests that
a specific collective process, independent of viscosity, may
operate and contribute in a way similar to that of mechanisms
present in viscous disks (see also Youdin \& Goodman
\cite{you05}). The very simple analytical study provided in the
present paper offers a description, at the linear level, of the
collective instabilities underlying such interesting process.

\begin{acknowledgements}
We wish to thank Giuseppe Lodato for a number of interesting
suggestions and Donald Lynden-Bell for useful conversations on the
subject of this paper. This work was partially supported by MIUR
of Italy (cofin-2004).
\end{acknowledgements}

\end{document}